\documentclass[prl,aps,twocolumn,superscriptaddress,nofootinbib,floatfix,
nolongbibliography]{revtex4-2}

\usepackage{amsmath,amssymb,bm}
\usepackage{graphicx}
\usepackage[T1]{fontenc}
\usepackage[hidelinks]{hyperref}
\newcommand{\bvec}[1]{\boldsymbol{#1}}
\newcommand{\modelplot}[1]{%
  \includegraphics[width=\columnwidth]{#1}}

\begin{document}

\title{Evading Sudakov Dilution in Gluon Tomography}

\author{Shuo Lin}
\email{shuolin@sdu.edu.cn}
\affiliation{Key Laboratory of Particle Physics and Particle Irradiation (MOE),
Institute of Frontier and Interdisciplinary Science, Shandong University,
Qingdao, Shandong 266237, China}

\author{Jian Zhou}
\email{jzhou@sdu.edu.cn}
\affiliation{Key Laboratory of Particle Physics and Particle Irradiation (MOE),
Institute of Frontier and Interdisciplinary Science, Shandong University,
Qingdao, Shandong 266237, China}
\affiliation{Southern Center for Nuclear-Science Theory (SCNT), Institute of
Modern Physics, Chinese Academy of Sciences, Huizhou, Guangdong 516000, China}

\begin{abstract}
Sudakov broadening and competing azimuthal harmonics from
final-state soft radiation limit the precision of gluon tomography.  We
suppress both effects using fiducial hadronic recoil---the vector sum of all
hadronic transverse momenta within a rapidity interval---while a tagged jet
fixes the azimuthal axis.  Momentum conservation ensures that emissions inside
the interval contribute neither to recoil broadening nor to these harmonics, so
widening the interval reduces their impact without a soft-radiation veto.  As a benchmark,
we study the $\cos 2\phi$ modulation
probing $h_1^{\perp g}$ in DIS.  For  rapidity interval  $\Delta\eta=3$, fiducial recoil enhances the
$h_1^{\perp g}$ contribution by a factor of $4.6$ and
suppresses the final-state soft-gluon term by a factor of $4.0$ relative to the
conventional dijet imbalance.  Fiducial recoil
thereby enables precision gluon-TMD tomography.
\end{abstract}

\maketitle

\textit{Introduction.---}
Mapping the transverse motion and polarization of gluons inside the proton is a
central goal of nucleon tomography.  Gluon transverse-momentum-dependent (TMD)
distributions provide this multidimensional picture.  A particularly revealing
example is $h_1^{\perp g}$, which correlates gluon transverse momentum with
linear polarization \cite{Mulders:2000sh}.  Linear gluon polarization probes
saturation dynamics and is tied to maximal spin--orbit entanglement in the
small-$x$ regime; in Higgs production, it leaves parity-sensitive imprints on
the transverse-momentum spectrum
\cite{Metz:2011wb,Dominguez:2011br,Bhattacharya:2024sno,Boer:2011kf}.
Most proposed measurements of $h_1^{\perp g}$ focus on dijet and
heavy-flavor production
\cite{Akcakaya:2012si,Lansberg:2015hla,Lansberg:2017dzg,Marquet:2017xwy,
Dumitru:2015gaa,Schafer:2012yx,Boer:2016fqd,Dumitru:2017ftq,Dumitru:2018kuw,Kishore:2018ugo,Efremov:2017iwh,Efremov:2018myn,
delCastillo:2021znl,Fucilla:2026mkg,Boer:2010zf,Boer:2011gc,Li:2023gkh,
Song:2025bdj,Altinoluk:2021ygv,Boussarie:2021ybe,Mantysaari:2019hkq}.
Despite substantial theoretical progress
\cite{Dominguez:2011wm,vanHameren:2021sqc,Gutierrez-Reyes:2019rug,
Zhu:2025ixc,Zhao:2022qym,Echevarria:2015uaa,Scarpa:2019fol,Bor:2022fga,Qiu:2011ai},
$h_1^{\perp g}$ remains unmeasured, and clean collider access to it remains
a central challenge.

Gluon-TMD measurements with colored final states face two distinct radiative
effects.  Initial-state radiation drives Sudakov broadening, while final-state
soft gluons generate additional azimuthal harmonics, including a competing
$\cos 2\phi$ modulation in measurements of $h_1^{\perp g}$
\cite{Boer:2009nc,Hatta:2020bgy,Hatta:2021jcd,delCastillo:2020omr}.  In the conventional dijet
observable, radiation outside the two reconstructed jets is omitted from the
recoil sum and therefore appears directly as transverse-momentum imbalance.
The resulting broadening shifts the spectrum toward larger imbalance and
weakens gluon-TMD sensitivity in the small-imbalance region.

\begin{figure}[!t]
 \centering
 \includegraphics[width=\columnwidth]{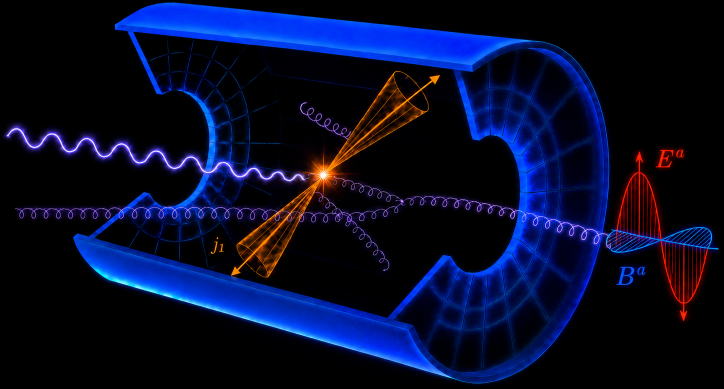}
 \caption{Schematic of the hadronic-recoil measurement in
 $\gamma^*g\to q\bar q$.  The tagged jet fixes the azimuthal reference axis.
 The recoil sum includes all hadronic transverse momenta inside the fiducial
 region.  The illustration was generated with GPT Image 2.}
 \label{fig:recoil-schematic}
\end{figure}

Ref.~\cite{Fang:2025dee} recently proposed a 0-jettiness requirement to
suppress central initial-state radiation and enhance TMD spin asymmetries, but
the accompanying soft-radiation veto reduces acceptance.  Rapidity-restricted
veto observables have also been studied in soft-collinear effective
theory~\cite{Hornig:2017pud,Michel:2018hui}.  In contrast, we include soft radiation
in the recoil rather than vetoing it.  The
hadronic recoil is the vector sum of all hadronic transverse momenta within a
fiducial interval.  Momentum conservation then ensures that radiation inside the
interval generates no recoil broadening: only radiation outside it contributes.
The same recoil definition therefore applies to other gluon-TMD observables.
Enlarging the interval reduces the out-of-interval phase space, weakens Sudakov
broadening, and preserves gluon-TMD sensitivity at small recoil.
For the $\cos 2\phi$ modulation, the competing final-state term is suppressed
even more strongly because the harmonic weight of soft emissions decreases
rapidly with their rapidity separation from the outgoing partons.  The interval
width thus controls both effects without a soft-radiation veto.
To implement the measurement, a single tagged jet fixes the azimuthal axis,
while lepton--jet kinematics determine whether the other hard parton lies inside
the interval
(Fig.~\ref{fig:recoil-schematic}).

We illustrate the gain with an EIC benchmark for $h_1^{\perp g}$, using the TMD
input specified below.  For a rapidity interval of
width $\Delta\eta=3$ and a recoil magnitude of $2\,{\rm GeV}$, the
$h_1^{\perp g}$ contribution to the $\cos 2\phi$ moment is enhanced
by a factor of $4.6$ relative to the $R=0.4$ dijet imbalance, while the
final-state soft-gluon contribution is suppressed by a factor of $4.0$.  As the
interval widens, the polarized-gluon contribution grows while the final-state
term decreases.  The azimuthally averaged channel likewise exhibits an
enhanced low-recoil yield.
Together, these gains establish fiducial hadronic recoil as a general strategy
for precision gluon-TMD tomography.

\textit{Theoretical formalism.---}
We consider the hard partonic process
\begin{equation*}
 \gamma^*(q)+g(x_gP)\to q(p_q)+\bar q(p_{\bar q}),
\end{equation*}
where $P$ is the proton momentum, $Q^2=-q^2$, and $x_g$ is the longitudinal
momentum fraction of the incoming gluon.  Bold symbols denote two-dimensional
transverse vectors.  A tagged jet with transverse momentum
$\bvec p_J$ and $P_T=|\bvec p_J|$ sets the azimuthal reference axis.  At parton
level, we identify this axis with the direction of the outgoing parton associated
with the tagged jet.  We define
$s_{\gamma p}=(P+q)^2$, $\hat s=(x_gP+q)^2$, and
$z=P\!\cdot p_q/(P\!\cdot q)$.  Two-body kinematics imply
$\hat s=P_T^2/[z(1-z)]$ and
$x_g=(Q^2+\hat s)/(Q^2+s_{\gamma p})$.  For the analytic presentation and
numerical benchmark, we take $z=1/2$.  The two hard outgoing partons then have
equal Breit-frame rapidities.  Throughout, $\eta$ is defined relative to this
common rapidity, so both partons are at $\eta=0$.

For a rapidity interval of width $\Delta\eta$ with full azimuthal coverage, the
hadronic recoil is the vector sum of the momenta of all hadrons inside it.  We denote the recoil by
$\bvec l$ and define $\phi$ as its azimuthal angle relative to the tagged-jet
momentum $\bvec p_J$:
\begin{equation}
 \bvec l=
 \sum_{|\eta_h|\leq\Delta\eta/2}\bvec p_h,\qquad
 l_T=|\bvec l|,\qquad
 \phi=\angle(\bvec l,\bvec p_J).
 \label{eq:fiducial-observable}
\end{equation}
We restrict the sample to the two-parton hard topology by excluding events with
an additional resolved hard jet.  This selection does not veto soft radiation
entering the fiducial recoil.  Within this topology, the small-recoil condition
$l_T\ll P_T$ requires both hard partons to lie inside the fiducial interval: if
the untagged parton falls outside, the tagged-parton transverse momentum remains
uncompensated in the measured sum, giving $l_T\sim P_T$.  We can also verify this
condition directly from lepton--jet kinematics.  Using the measured lepton and
tagged-jet momenta, we fix $x_g$ through $(x_gP+q-p_J)^2=0$ and reconstruct the
untagged-parton momentum as $x_gP+q-p_J$, which allows us to determine whether
it lies inside the interval.  The jet radius $R$ affects only the tagged-jet
selection, not the recoil definition.

The active incoming gluon carries transverse momentum $\bvec\kappa$.  With full
final-state coverage, momentum conservation gives $\bvec l=\bvec\kappa$.  The
fiducial sum omits radiation outside the interval, so the measured recoil becomes
\begin{equation}
 \bvec l
 =\bvec\kappa-
 \sum_{a\in{\rm out}}\bvec k_a .
 \label{eq:measurement-function}
\end{equation}
Here $\bvec k_a$ is the transverse momentum of emission $a$, and the sum includes
all emissions with relative rapidity $|\eta_a|>\Delta\eta/2$.  Radiation inside
the interval remains in the measured sum and therefore causes no additional
recoil broadening.

At Born level, the recoil dependence is governed by the unpolarized and linearly
polarized gluon TMDs, $f_1^g$ and $h_1^{\perp g}$.  For compactness, we define
$h_g(x_g,l_T)\equiv[l_T^2/(2M_p^2)]h_1^{\perp g}(x_g,l_T)$, where $M_p$ is the
proton mass, and suppress the common arguments $(x_g,l_T)$ below.  The cross
sections for longitudinal and transverse photons are\cite{Metz:2011wb,Dominguez:2011br,Pisano:2013cya,
Dumitru:2018kuw}.
\begin{align}
&d\sigma_{L,T}^{\rm Born}
=
\sigma_0^{L,T}x_g\!\left[
f_1^g+2a_{L,T}h_g\cos(2\phi)\right],
 \label{eq:angular-coefficients}
\\
&a_L=\frac12,\qquad a_T=
-\frac{z(1-z)Q^2P_T^2}
{[z(1-z)Q^2]^2+P_T^4}.
\end{align}
The prefactors $\sigma_0^{L,T}$ contain the hard-scattering and phase-space
factors.

Beyond Born level, soft radiation from the colored hard legs modifies both the
recoil spectrum and its azimuthal structure\cite{Catani:2014qha,Catani:2017tuc,Kang:2020xez,Fu:2026nkd,Kang:2021ffh,Chien:2022wiq}.  For the fiducial recoil, however,
only emissions outside the interval change the measured recoil.  Following the
standard eikonal treatment of single-soft-gluon emission, we multiply the usual
eikonal kernel by $k_T^2$ to define the dimensionless angular antenna:
\begin{equation}
 \mathcal A(k)=\sum_{i<j}(-\boldsymbol T_i\!\cdot\!\boldsymbol T_j)
 \frac{k_T^2\,p_i\!\cdot p_j}{(p_i\!\cdot k)(p_j\!\cdot k)} .
 \label{eq:eikonal-antenna}
\end{equation}
With this convention, the radial emission measure is $dk_T/k_T$.
The indices $i,j\in\{g,q,\bar q\}$ label the three colored hard legs, whose
momenta and color charges are $p_i$ and $\boldsymbol T_i$.  The emitted gluon
carries momentum $k$, with $k_T=|\bvec k|$.  Dots between color charges denote
color-space products; all other dots denote Lorentz products.  For the symmetric
configuration $z=1/2$ adopted here, the antenna depends only on the relative
rapidity $\eta$ and azimuth $\varphi$ of the emission.  Its relevant angular
projections are
\begin{equation}
 \mathcal A_n(\eta)\equiv
 \int_0^{2\pi}\frac{d\varphi}{2\pi}\cos(n\varphi)
 \mathcal A(\eta,\varphi),\qquad n=0,2 .
 \label{eq:antenna-moments}
\end{equation}
The $n=0$ projection controls recoil broadening, while the $n=2$ projection
isolates the soft-gluon contribution to the $\cos 2\phi$ harmonic.  The
measurement function in Eq.~\eqref{eq:measurement-function} restricts both
contributions to radiation outside the interval.  At $z=1/2$,
$\mathcal A_2(\eta)=e^{-2|\eta|}\mathcal A_0(\eta)$, so widening the fiducial
interval exponentially suppresses this out-of-interval harmonic.

To resum the logarithmically enhanced recoil broadening, we perform the
calculation in impact-parameter $\bvec b$ space, where $\bvec b$ is Fourier conjugate
to $\bvec l$.  The Fourier transformation converts transverse-momentum
convolutions into products.  With the phase convention
$e^{-i\bvec b\cdot\bvec l}$, the measurement weight is unity for emissions
inside the interval and $e^{i\bvec b\cdot\bvec k}$ for emissions outside it.
The real contribution from inside the interval therefore cancels the
corresponding virtual correction and causes no recoil broadening.  After
azimuthal integration, the real--virtual combination gives $1-J_0(bk_T)$ for
$\mathcal A_0$.  For $\mathcal A_2$, the virtual correction drops out because
it is azimuthally symmetric.  The remaining real-emission weight is
$J_2(bk_T)$.

The two projections are treated differently.  The $k_T$ integral of
$\mathcal A_0$ generates the initial-state cusp logarithm and a single logarithm
from radiation outside the interval.  These logarithms must be resummed.  By
contrast, $\mathcal A_2$ gives the
final-state soft-gluon contribution to the $\cos 2\phi$ modulation, which we
retain at one-emission accuracy.  As $\Delta\eta$ increases, the upper $k_T$
limits for out-of-interval radiation decrease exponentially, weakening the
Sudakov broadening.

Following the standard resummation procedure of
Refs.~\cite{Mueller:2012uf,Mueller:2013wwa,Zhou:2016tfe,Xiao:2017yya,Zhou:2018lfq,Zheng:2019zul,Sun:2014gfa,Sun:2015doa,Liu:2020dct,Hatta:2019ixj,Caucal:2022ulg,Caucal:2023nci,Caucal:2023fsf,Shao:2026doo,Liu:2018trl},
the logarithms generated by $\mathcal A_0$ exponentiate into a Sudakov factor.
The evolution runs from the canonical $b$-space scale $\mu_b$ to the hard scale
$P_T$.  Its kernel contains the standard gluon noncusp term $-2C_A\beta_0$ and
a finite out-of-interval contribution $c_{\rm fid}(\Delta\eta)$.  The latter is
analogous to the jet-radius coefficient $c_0(R)$ for the dijet imbalance.
Including the common nonperturbative contribution $S_{\rm NP}$, the exponent is
\begin{align}
&S(b;\Delta\eta)
=S_{\rm NP}(b)
+\int_{\mu_b}^{P_T}\frac{d\mu}{\mu}
\frac{\alpha_s(\mu)C_A}{\pi}
\label{eq:recoil-exponent}
\\
&\quad\times\Biggl[
-2\beta_0+\ln\frac{P_T^2}{\mu^2}
+2\ln\frac{\hat s+Q^2}{\hat s}
+\frac{2C_F}{C_A}c_{\rm fid}(\Delta\eta)
\Biggr].\notag
\end{align}
Here $\beta_0$ is the one-loop beta-function coefficient.  The same exponent
applies to the unpolarized and linearly polarized gluon terms.

The resummed distribution is obtained by Fourier inversion, with $J_0$ for the
azimuthal average and $J_2$ for the second harmonic.  We write these transforms
as
\begin{equation}
 \mathcal B_n[F]
 \equiv\frac{1}{2\pi}\int_0^\infty db\,bJ_n(bl_T)F(b).
 \label{eq:Bn}
\end{equation}
The unpolarized input is the collinear gluon density $f_g(x_g,\mu_b)$, while the
linearly polarized input is the $b$-space coefficient $h_g(x_g,b)$.  The
resummed angular distributions are then
\begin{align}
 d\sigma_{L,T}^{\rm resum}={}&\sigma_0^{L,T}x_g
 \mathcal B_0[e^{-S}f_g]\notag
 \\
&\times[1+2\langle\cos2\phi\rangle_{L,T}\cos(2\phi)].
 \end{align}

The second-harmonic moments are defined separately for longitudinal and
transverse photons.  For each polarization, the $J_2$ coefficient is normalized
to the corresponding azimuthally averaged $J_0$ spectrum.  The numerator
contains both the linearly polarized gluon term and the final-state soft-gluon
contribution:
\begin{align}
 \langle\cos2&\phi\rangle_{L,T}=
 \frac{1}{\mathcal B_0[e^{-S}f_g]}
 \mathcal B_2\Bigl[e^{-S}\Bigl\{
 a_{L,T}h_g\notag
 \\[-0.2ex]
 &\hspace{-0.5em}{}+f_g\frac{\alpha_s(\mu_b)}{\pi}
 \int_{\rm out}\frac{dk_T}{k_T}\,d\eta\,
 J_2(bk_T)\mathcal A_2(\eta)
 \Bigr\}\Bigr] .
 \label{eq:resummed-moment}
\end{align}
Here, ``out'' denotes the allowed phase space outside the fiducial interval, and
the $k_T$ integral extends to the rapidity-dependent upper limit specified in the
Supplemental Material.  The final-state soft-gluon term is independent of photon
polarization and therefore enters the longitudinal and transverse moments with
the same coefficient.

\textit{Phenomenology.---}
We quantify the gain from fiducial hadronic recoil by comparing it with the
conventional dijet imbalance at the same EIC kinematics \cite{AbdulKhalek:2021gbh}.  We choose
$Q=10\,{\rm GeV}$,
$P_T=15\,{\rm GeV}$,
$z=\tfrac12$, and
$\sqrt{s_{\gamma p}}=100\,{\rm GeV}$, corresponding to $x_g\simeq0.1$.
The collinear gluon density and $\alpha_s$ are taken from the central
HERAPDF20 NLO set \cite{H1:2015ubc}.  We
model $h_1^{\perp g}$ by combining the Gaussian phenomenological input of
Ref.~\cite{Hatta:2021jcd} with the one-loop TMD matching contribution
\cite{Gutierrez-Reyes:2019rug}.  The same $f_g$ and $h_1^{\perp g}$ inputs are used
for both observables; full prescriptions are given in the Supplemental
Material.  We evaluate the hadronic recoil for
$\Delta\eta=1,2,3$.

For jet transverse momenta $\bvec p_1$ and $\bvec p_2$,
the dijet imbalance is $\bvec p_1+\bvec p_2$, and its azimuth is measured from
$\bvec p_1-\bvec p_2$.  At $R=0.4$, we use $c_0=3.14$ and $c_2=0.96$ from
Ref.~\cite{Hatta:2021jcd}.  For the dijet benchmark, the Sudakov exponent also
includes the nonperturbative contribution from the final-state jets.  We set
$c_{\rm fid}=c_0$ in Eq.~\eqref{eq:recoil-exponent} and replace the $\mathcal A_2$
integral by $C_Fc_2$ in Eq.~\eqref{eq:resummed-moment}.

\begin{figure}[!htbp]
 \modelplot{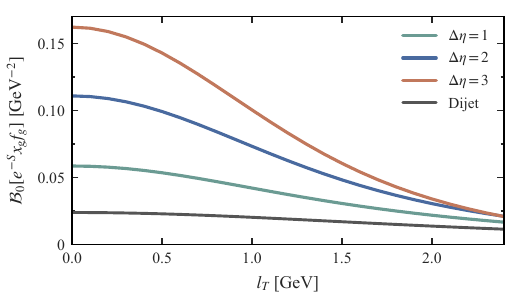}
 \caption{Azimuthally averaged hadronic-recoil spectra for
 $\Delta\eta=1,2,3$, compared with the $R=0.4$ dijet-imbalance spectrum at the
 same hard kinematics and with common TMD inputs.  Both hard outgoing partons
 lie inside the fiducial interval.  For either observable, $l_T$ denotes the
 corresponding recoil magnitude; the common hard prefactor has been divided
 out.}
 \label{fig:unpolarized-comparison}
\end{figure}

At low $l_T$, the hadronic-recoil spectrum exceeds the $R=0.4$ dijet result
(Fig.~\ref{fig:unpolarized-comparison}).  The excess becomes more pronounced as
$\Delta\eta$ increases.  Weaker Sudakov broadening narrows the recoil spectrum,
shifting events from larger to smaller $l_T$ while leaving the
recoil-integrated rate unchanged in our resummed treatment.  The larger
low-$l_T$ yield improves sensitivity to gluon TMDs.
\begin{figure}[!htbp]
 \modelplot{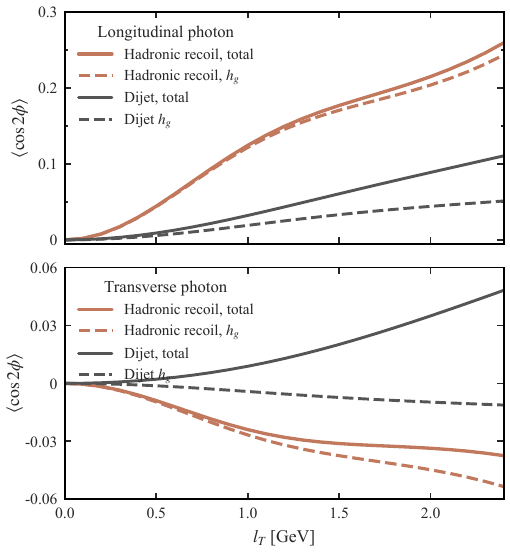}
 \caption{$\cos2\phi$ moments for the hadronic recoil at $\Delta\eta=3$ and the
 $R=0.4$ dijet imbalance at the chosen kinematics.  Compared with the dijet
 imbalance, the hadronic recoil enhances the linearly polarized-gluon
 contribution and suppresses the final-state soft-gluon contribution.  Solid
 curves show the full moments, and dashed curves show the linearly
 polarized-gluon contributions; their difference is the final-state soft-gluon
 contribution.}
 \label{fig:dijet-comparison}
\end{figure}

\begin{figure}[!htbp]
 \modelplot{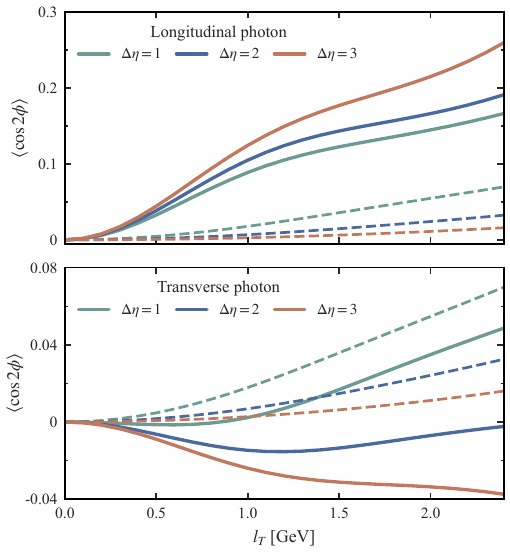}
 \caption{Hadronic-recoil $\cos2\phi$ moments for $\Delta\eta=1,2,3$ at the
 chosen kinematics.  As the interval widens, the magnitude of the linearly
 polarized-gluon contribution grows while the final-state soft-gluon
 contribution decreases.  Solid curves show the full results; dashed curves
 show the final-state soft-gluon contribution from radiation outside the
 interval.}
 \label{fig:span-dependence}
\end{figure}

At $\Delta\eta=3$ and $l_T=2\,{\rm GeV}$, the linearly polarized-gluon
contribution to $\langle\cos 2\phi\rangle_L$ is $4.6$ times its value for the
dijet imbalance.  The final-state soft-gluon contribution is smaller by a
factor of $4.0$ (Fig.~\ref{fig:dijet-comparison}).  Compared with the dijet
imbalance, fiducial recoil enhances both the low-$l_T$
yield and the linearly polarized-gluon contribution.  In both cases, the
enhancement results from weaker Sudakov suppression because only out-of-interval
radiation broadens the recoil.  The additional reduction of the final-state term
follows from $\mathcal A_2(\eta)=e^{-2|\eta|}\mathcal A_0(\eta)$, which
exponentially suppresses the harmonic weight at large rapidity separation from
the outgoing partons.

For longitudinal photons, the polarized-gluon and final-state soft-gluon terms
have the same sign.  For transverse photons, $a_T<0$ makes the polarized-gluon
term negative while the soft-gluon term remains positive.  The polarized-gluon
term dominates the hadronic-recoil moment, giving $\langle\cos2\phi\rangle_T<0$,
whereas the soft-gluon term dominates the dijet-imbalance moment,
giving
$\langle\cos2\phi\rangle_T>0$.  The polarization-independent final-state term
cancels in $\langle\cos2\phi\rangle_L-\langle\cos2\phi\rangle_T$, so this
difference isolates the $h_1^{\perp g}$ contribution.  At $l_T=2\,{\rm GeV}$,
this difference is $0.25$ for the hadronic recoil, a factor of $4.6$
larger than for the dijet imbalance.  At these kinematics, the phenomenological
and perturbative components of $h_1^{\perp g}$ make comparable contributions.  The
magnitudes and relative sign of the hadronic-recoil and dijet transverse-photon
moments therefore depend on the assumed input.

As the interval widens, the linearly polarized-gluon contribution grows in
magnitude while the final-state soft-gluon contamination decreases
(Fig.~\ref{fig:span-dependence}).  In the transverse-photon channel, the soft
term dominates at $\Delta\eta=1$, while the two contributions are comparable
at $\Delta\eta=2$.  At $\Delta\eta=3$, the polarized-gluon contribution
dominates both photon polarizations, making the hadronic-recoil modulation a
clean probe of $h_1^{\perp g}$ without a soft-radiation veto.

\textit{Conclusion.---}
We propose fiducial hadronic recoil as a novel strategy for gluon TMD
tomography.  It simultaneously suppresses Sudakov broadening and final-state
soft-gluon contamination without vetoing radiation or sacrificing event
statistics.  Compared with the conventional dijet imbalance, fiducial recoil
yields substantially more low-$l_T$ events and a much cleaner $h_1^{\perp g}$
modulation, improving sensitivity to nonperturbative gluon motion and
spin--momentum correlations.

Experimentally, a single tagged jet defines the transverse reference axis, and
the recoil is the vector sum of all
hadrons in the fiducial interval.  For the unpolarized gluon TMD, lepton--jet
kinematics need only verify that the other hard parton lies inside the interval;
no second-jet tag or precise reconstruction is required.  Because the
momentum-conservation cancellation is spin independent, fiducial recoil can be
extended to other leading-twist gluon TMDs through appropriate beam and
target polarizations and azimuthal projections.  Fiducial recoil thus opens a
new avenue for precision nucleon tomography.

\noindent{\it Acknowledgments.}
This work was supported by the National Natural Science Foundation of China under Grant
Nos.~12175118, 12321005 and 12575091 (J.Z.), and Project SDCX-ZG-202603015 funded by Postdoctoral Innovation Program of Shandong Province(S.L).

\bibliography{ref}

\clearpage
\onecolumngrid
\appendix
\setcounter{equation}{0}
\renewcommand{\theequation}{S\arabic{equation}}
\renewcommand{\thefigure}{S\arabic{figure}}
\renewcommand{\theHfigure}{S\arabic{figure}}
\setcounter{figure}{0}
\section*{Supplemental Material}

This Supplemental Material specifies the TMD inputs and dijet benchmark used in
the numerical analysis and derives the contribution of soft radiation outside
the fiducial rapidity interval.  Out-of-interval radiation determines both the
finite Sudakov coefficient $c_{\rm fid}(\Delta\eta)$ and the final-state
$\cos 2\phi$ harmonic used in the phenomenology.

\section{TMD input}
\label{app:tmd}

The momentum-space gluon correlator is
\begin{equation}
 {\cal M}^{ij}(x,\bvec\kappa)=\frac{x}{2}\left[
 \delta^{ij}f_1^g(x,\kappa_T)
 +\left(2\frac{\kappa^i\kappa^j}{\kappa_T^2}-\delta^{ij}\right)
 \frac{\kappa_T^2}{2M_p^2}h_1^{\perp g}(x,\kappa_T)\right] .
 \label{eq:gluon-correlator}
\end{equation}
As in Eq.~\eqref{eq:angular-coefficients}, we denote the coefficient of the
traceless tensor by $h_g(x,\kappa_T)$.  With Fourier phase
$e^{i\bvec b\cdot\bvec\kappa}$, the traceless tensor transforms as
\begin{equation}
 \int_0^{2\pi}\frac{d\phi_\kappa}{2\pi}
 e^{i\bvec b\cdot\bvec\kappa}
 \left(2\frac{\kappa^i\kappa^j}{\kappa_T^2}-\delta^{ij}\right)
 =-J_2(b\kappa_T)
 \left(2\frac{b^ib^j}{b^2}-\delta^{ij}\right).
 \label{eq:supp-fourier}
\end{equation}
In $\widetilde{\cal M}^{ij}/x$, $h_g(x,b)$ is the scalar coefficient of the tensor
$-\tfrac12(2b^ib^j/b^2-\delta^{ij})$.  This convention absorbs the minus sign in
Eq.~\eqref{eq:supp-fourier}.  The corresponding inverse transform is
\begin{equation}
 h_g(x,\kappa_T)=
 \int_0^\infty \frac{b\,db}{2\pi}\,
 J_2(\kappa_T b)h_g(x,b).
 \label{eq:supp-hg-transform}
\end{equation}
It therefore has the positive $J_2$ kernel used in $\mathcal B_2$.

The $b_*$ and scale prescriptions are
\begin{equation}
 b_*=\frac{b}{\sqrt{1+b^2/(1.5\,{\rm GeV}^{-1})^2}},\qquad
 \mu_b=\min\!\left[P_T,\max\!\left(
 1\,{\rm GeV},\frac{2e^{-\gamma_E}}{b_*}\right)\right].
 \label{eq:supp-bstar}
\end{equation}
The bound $\mu_b\geq1\,{\rm GeV}$ keeps the PDF and coupling evaluations within
their scale grids.  We take $\alpha_s$ and the unpolarized input
$f_g(x_g,\mu_b)$ from member $0$ of the HERAPDF20 NLO set.  Following
Ref.~\cite{Hatta:2021jcd}, we model $h_g(x_g,b)$ as the sum of a phenomenological
term with Gaussian transverse-momentum dependence and a perturbative term from
one-loop matching:
\begin{align}
 h_g(x_g,b)={}&f_g(x_g,\mu_b)
 \frac{ eQ_h^2b^2}{27}
 \exp\!\left(\frac{Q_h^2b^2}{12}\right)
 \notag\\
 &+\frac{\alpha_s(\mu_b)}{\pi x_g}
 \int_{x_g}^1\frac{dx'}{x'}(x'-x_g)
 \left\{C_A f_g(x',\mu_b)
 +C_F\sum_q\left[q(x',\mu_b)+\bar q(x',\mu_b)\right]\right\} .
 \label{eq:supp-tmd-inputs}
\end{align}
Here $Q_h=1\,{\rm GeV}$.  The first term in
Eq.~\eqref{eq:supp-tmd-inputs} follows the ansatz of
Ref.~\cite{Boer:2011kf}.  The second term is the one-loop matching contribution
summed over five quark flavors.
The nonperturbative exponent for the fiducial recoil is
\begin{equation}
 S_{\rm NP}(b)=\frac{C_A}{C_F}\left[
 0.106\,{\rm GeV}^2 b^2
 +0.42\ln\frac{P_T}{Q_0}\ln\frac{b}{b_*}\right] .
 \label{eq:supp-SNP}
\end{equation}
Here $Q_0^2=2.4\,{\rm GeV}^2$.  The factor $C_A/C_F$ in
Eq.~\eqref{eq:supp-SNP} implements Casimir scaling of the initial-state
nonperturbative contribution from quarks to gluons.

\section{Soft radiation outside the rapidity interval}
\label{app:soft-factor}

At $z=1/2$, the two outgoing partons have the common Breit-frame rapidity
$\ln(2P_T/Q)$.  Throughout this section, $\eta$ denotes rapidity measured
relative to this value.  The logarithmically enhanced phase space satisfies
$k_Te^{-\eta}<P_T$ and $k_Te^\eta<Q_+$, where
$Q_+=(\hat s+Q^2)/(2P_T)$.  Outside the fiducial interval, these bounds restrict
$\eta$ to
\begin{equation}
 -\ln\frac{P_T}{k_T}<\eta<-\frac{\Delta\eta}{2},\qquad
 \frac{\Delta\eta}{2}<\eta<\ln\frac{Q_+}{k_T}.
 \label{eq:supp-rapidity-support}
\end{equation}
The negative- and positive-rapidity ranges are nonempty for
$k_T<P_Te^{-\Delta\eta/2}$ and $k_T<Q_+e^{-\Delta\eta/2}$, respectively.

Color conservation gives a color factor $C_A/2$ for each incoming--outgoing
dipole and $(2C_F-C_A)/2$ for the outgoing dipole.  Projecting the eikonal
antenna in Eq.~\eqref{eq:eikonal-antenna} at $z=1/2$ yields
\begin{align}
 \mathcal A_0(\eta)={}&2\bigl(C_Ae^\eta\cosh\eta+2C_F-C_A\bigr)
 \operatorname{csch}(2|\eta|),
 \notag\\[-0.1ex]
 \mathcal A_2(\eta)={}&e^{-2|\eta|}\mathcal A_0(\eta).
 \label{eq:supp-antenna-moments}
\end{align}
The final-state soft-gluon contribution in
Eq.~\eqref{eq:resummed-moment} is obtained by inserting $\mathcal A_2$ from
Eq.~\eqref{eq:supp-antenna-moments} and integrating over the rapidity support in
Eq.~\eqref{eq:supp-rapidity-support}.

Integrating $\mathcal A_0$ over the rapidity ranges in
Eq.~\eqref{eq:supp-rapidity-support} and keeping the terms that survive as
$k_T\to0$ gives
\begin{equation}
 \int_{\rm out}d\eta\,\mathcal A_0
 =2C_A\ln\frac{Q_+e^{-\Delta\eta/2}}{k_T}
 +2(4C_F-C_A)\operatorname{atanh}(e^{-\Delta\eta})
 -C_A\ln(1-e^{-2\Delta\eta})
 +\mathcal O\!\left(\frac{k_T^2}{P_T^2},
 \frac{k_T^2}{Q_+^2}\right).
 \label{eq:supp-A0-limit}
\end{equation}
To match the normalization used for the DIS dijet coefficient $c_0(R)$, we
define the azimuthally averaged coefficient as
\begin{equation}
 c_{\rm fid}(\Delta\eta)\equiv\frac{1}{2C_F}\lim_{k_T\to0}\left[
 \int_{\rm out}d\eta\,\mathcal A_0
 -C_A\ln\frac{P_T^2}{k_T^2}
 -2C_A\ln\frac{\hat s+Q^2}{\hat s}\right].
 \label{eq:supp-c0det-definition}
\end{equation}
At $z=1/2$, Eqs.~\eqref{eq:supp-A0-limit} and
\eqref{eq:supp-c0det-definition} give
\begin{equation}
 c_{\rm fid}(\Delta\eta)=\frac{1}{2C_F}\left[
 2(4C_F-C_A)\operatorname{atanh}(e^{-\Delta\eta})
 -C_A\ln(1-e^{-2\Delta\eta})+C_A(2\ln2-\Delta\eta)\right].
 \label{eq:supp-c0det}
\end{equation}
For $\Delta\eta=1,2,3$, this gives
$c_{\rm fid}=1.27$, $-0.431$, and $-1.73$, respectively.
For five active flavors, $\beta_0=11/12-5/(6C_A)$.  In the Sudakov exponent of
Eq.~\eqref{eq:recoil-exponent}, $c_{\rm fid}(\Delta\eta)$ from
Eq.~\eqref{eq:supp-c0det} replaces the DIS dijet coefficient $c_0(R)$.
In the numerical evaluation, the perturbative kernel in square brackets in
Eq.~\eqref{eq:recoil-exponent} can become negative over part of the $\mu$
integration range.  We set it to zero in that region instead of extrapolating
the truncated logarithmic approximation beyond its range of validity.

For comparison, the dijet reference includes an additional nonperturbative
term $2g_\Lambda b^2$ from the two final-state jets in the model of
Ref.~\cite{Hatta:2021jcd}.  We use $g_\Lambda=0.1\,{\rm GeV}^2$.  The resulting
Sudakov exponent is
\begin{align}
 S_{\rm dijet}(b)={}&S_{\rm NP}(b)+2g_\Lambda b^2
 \notag\\
 &+\int_{\mu_b}^{P_T}\frac{d\mu}{\mu}
 \frac{\alpha_s(\mu)C_A}{\pi}
 \left[-2\beta_0+\ln\frac{P_T^2}{\mu^2}
 +2\ln\frac{\hat s+Q^2}{\hat s}
 +\frac{2C_F}{C_A}c_0\right],
 \label{eq:supp-dijet-exponent}
\end{align}
and the second-harmonic moment is
\begin{equation}
 \langle\cos2\phi\rangle_{L,T}^{\rm dijet}=
 \frac{\mathcal B_2\!\left[e^{-S_{\rm dijet}}
 \left\{a_{L,T}h_g(x_g,b)
 +f_g(x_g,\mu_b)\frac{\alpha_s(\mu_b)C_F}{\pi}c_2\right\}\right]}
 {\mathcal B_0\!\left[e^{-S_{\rm dijet}}f_g(x_g,\mu_b)\right]}.
 \label{eq:supp-dijet-moment}
\end{equation}
Thus $c_0$ controls the recoil broadening, whereas $c_2$ controls the
final-state $\cos2\phi$ harmonic.

\end{document}